\documentclass[11pt]{article} 

\usepackage[utf8]{inputenc}
\usepackage[english]{babel}
\usepackage{graphicx}
\usepackage[a4paper,right=1in,left=1in,top=1in,bottom=1in]{geometry}
\usepackage{fancyhdr}

\usepackage[dvipsnames]{xcolor}
\usepackage[round]{natbib}
\usepackage{xcolor,colortbl}
\definecolor{light-gray}{gray}{0.9}

\usepackage{amsmath, amsfonts, amssymb,mathabx} 
\usepackage[labelfont=bf]{caption}
\usepackage{setspace}
\usepackage{accents} 
\usepackage{color}
\usepackage[title,titletoc,toc]{appendix}
\usepackage{rotating} 
\usepackage{float} 
\usepackage{multirow} 
\usepackage[position=top]{subfig}
\usepackage[subfigure]{tocloft}
\usepackage{tikz}
\usetikzlibrary{decorations.pathreplacing,calc}
\usepackage[many]{tcolorbox}
\newtcolorbox{rightbrace}{%
    enhanced jigsaw, 
    breakable, 
    frame hidden, 
    parbox=false,
}
\usepackage[explicit]{titlesec}
\graphicspath{ {images/} }
\title{A stacked approach for chained equations multiple imputation incorporating the substantive model }
\author{\textbf{Lauren J. Beesley$^{*1}$ and Jeremy M G Taylor$^{1}$} \\
$^{1}$University of Michigan, Department of Biostatistics\\
*Corresponding Author: lbeesley@umich.edu
}
\date{}   

\allowdisplaybreaks
\usepackage{color}   
\usepackage{hyperref}
\hypersetup{
    colorlinks=true, 
    linktoc=all,     
    linkcolor=black,  
    filecolor=black,
    citecolor = black,      
    urlcolor=black,
}

\usepackage{xr}
\begin{document}
\maketitle

\abovedisplayskip=6pt
\belowdisplayskip=6pt
\allowdisplaybreaks
\raggedbottom

\begin{abstract}{
\indent Multiple imputation by chained equations (MICE) has emerged as a popular approach for handling missing data. A central challenge for applying MICE is determining how to incorporate outcome information into covariate imputation models, particularly for complicated outcomes. Often, we have a particular \textit{analysis model} in mind, and we would like to ensure congeniality between the imputation and analysis models. \\
\indent We propose a novel strategy for directly incorporating the analysis model into the handling of missing data. In our proposed approach, multiple imputations of missing covariates are obtained \textit{without using outcome information}. We then utilize the strategy of imputation stacking, where multiple imputations are stacked on top of each other to create a large dataset. The analysis model is then incorporated through weights. Instead of applying Rubin's combining rules, we obtain parameter estimates by fitting a \textit{weighted} version of the analysis model on the stacked dataset. We propose a novel estimator for obtaining standard errors for this stacked and weighted analysis. Our estimator is based on the observed data information principle in Louis (1982) and can be applied for analyzing stacked multiple imputations more generally. Our approach for analyzing stacked multiple imputations is the first well-motivated method that can be easily applied for a wide variety of standard analysis models and missing data settings. \\
\indent In simulations, the proposed strategy produced unbiased parameter estimates when the analysis model was correctly specified. We developed an R package, \textit{StackImpute}, allowing this imputation approach to be easily implemented for many standard analysis models. 
 }\end{abstract}



\section{Introduction}
\indent Missing data is a common problem in modern observational data analysis, and the handling and treatment of these missing data can often have a large impact on statistical inference \citep{Little2002}. In response, a suite of statistical methods has been developed to tackle the various challenges that arise. In particular, a statistical strategy called multiple imputation has emerged as a popular and attractive approach for handling missing data in a wide variety of settings. Under multiple imputation, we use statistical models to draw multiple versions of the missing data, resulting in $M$ complete datasets. Then, the desired analysis is applied to each complete dataset separately and combined across datasets using Rubin's combining rules \citep{Little2002}. The central challenge of multiple imputation is specifying the statistical models or distributions used to obtain the draws of the missing data.   \\
\indent Traditional multiple imputation strategies involve filling in values for the missing data by drawing from distributions obtained from an assumed \textit{joint distribution} for all the variables of interest. Rather than specifying a \textit{joint model} for all the variables of interest, an alternative strategy called multiple imputation by chained equations (MICE) involves specifying \textit{conditional distributions} for each variable with missingness directly \citep{Raghu2001, VanBuuren2006}. These imputation distributions can be very flexible (e.g. random forests), or they can be based on standard regression models. Generally, these imputation models will \textit{not} correspond to a valid joint distribution. Compared to imputation using a valid joint distribution, MICE has fewer theoretical guarantees \citep{Liu2013b, Hughes2014}. However, MICE is often easy to implement and understand, and it can accommodate complicated variable relationships such as bounds, nonlinearity, and interactions. Software development has made MICE readily accessible to analysts, leading MICE to become an essential tool in the statistical toolbox for handling missing data.\\
\indent With easy-to-use software at an analyst's fingertips, it can become tempting to throw MICE at any missing data problem without careful thought about the imputation distributions. Suppose our ultimate goal is to model the relationship between some outcome, $Y$, and covariates $X$. Suppose we have missingness in $X$ and possibly also in $Y$. Literature suggests that we should somehow incorporate information in $Y$ into the distributions used to impute missing values in $X$ \citep{Moons2006}. A particularly tricky problem arises when $Y$ is complicated. $Y$ may be a longitudinal or survival-type outcome, or the relationship between $Y$ and $X$ may be involve interactions. Incorporating complicated $Y$ into imputation models for $X$ can be challenging and can potentially have a large impact in terms of bias in downstream analyses \citep{Beesley2016}. \\
\indent \citet{Bartlett2014} proposes a strategy called SMC-FCS (substantive model compatible fully conditional specification) that uses the assumed $Y \vert X$ relationship directly to incorporate $Y$ into the imputation distributions. In particular, missing covariate $X^p$ is imputed from a distribution proportional to the outcome model $f(Y \vert X)$ multiplied by an assumed relationship between $X^p$ and the other covariates, $X^{-p}$. An advantage of this approach over traditional MICE is that the assumed relationship between $Y$ and $X$ used for \textit{imputation} is consistent with the assumed relationship in the \textit{analysis model}, called congeniality \citep{Meng1994}. A lack of congeniality can sometimes produce bias in the downstream analysis \citep{Robins2000}. Additionally, this imputation strategy can substantially simplify the task of incorporating $Y$ into the imputation of missing $X$. However, the resulting imputation distribution is often known only up to proportionality, and more advanced methods such as rejection sampling or Metropolis-Hastings methods must often be used to obtain imputed values for each $X^p$. An R package \textit{smcfcs} exists for implementing SMC-FCS in certain outcome modeling settings, but this method can require additional work to implement in general. \\
\indent In this paper, we propose a novel strategy for incorporating the outcome model structure into the imputation pipeline that maintains the advantages of the method in \citet{Bartlett2014} but is more easily implemented, particularly for complicated or non-standard $Y \vert X$. We utilize the strategy of imputation stacking, where multiple imputations of the missing data are stacked on top of each other to create a large dataset \citep{Robins2000, VanBuuren2018}. In our proposed approach, multiple imputations of missing $X$ are obtained using imputation distributions that \textit{do not involve the outcome $Y$}. While this approach will generally result in bias for \textit{standard} multiple imputation, our method attains \textit{valid parameter estimates} by augmenting the stacked dataset with weights defined using the $Y \vert X$ model structure. We then estimate parameters in the analysis model by fitting a \textit{weighted} model for $Y \vert X$ on the stacked dataset. This strategy allows imputation and data analysis to be easily performed by separate analysts without concerns about uncongeniality between the imputation and analysis models and the potential negative impact on inference. Additionally, this imputation stacking strategy is particularly useful in settings where we want to impose restrictions \textit{across} imputed datasets such as when variable selection is of primary interest \citep{Wood2008}. This work is the first to propose a statistical strategy for chained equations imputation that (1) directly incorporates the outcome model structure \textit{and} (2) involves imputation from standard models such as regression models.    \\
\indent While imputation stacking can produce valid parameter estimates when the imputation models are well-specified, additional work is needed to obtain valid standard error estimates \citep{Robins2000, VanBuuren2018}. \citet{Robins2000} and \citet{Kim2011} provide strategies for estimating standard errors using stacked, imputed data. As we will discuss later on, both approaches have substantial limitations that may reduce their usage in practice. \citet{Wood2008} proposes an approach for estimating standard errors that is easy to implement but weakly justified in settings where missingness is not completely random. In this paper, we develop an alternative strategy for estimating standard errors for data analysis using stacked multiple imputations, and this estimator can be applied in general imputation settings. Our approach for estimating standard errors based on stacked multiple imputations is the first proposed method that can be easily and routinely applied for a wide variety of standard analysis models and missing data settings. In particular, we have developed an accompanying R package \textit{StackImpute} that will allow the proposed estimation to be easily implemented for many popular regression models including generalized linear models and Cox proportional hazards models. \\
\indent In Section 2 of this paper, we detail our proposed imputation algorithm and its theoretical motivation. In Section 3, we provide a strategy for estimating standard errors. In Section 4, we demonstrate the potential of our proposed method through a simulation study. In Section 5, we apply this imputation approach to handle missing data in a study of overall survival and time to recurrence for patients with head and neck cancer. In Section 6, we present a discussion.


\section{Imputation Strategy} \label{imp}
Suppose we are interested in the relationship between outcome $Y$ and covariate variables represented by matrix $X$. We will assume for now that $Y$ is fully observed, and we will extend to the setting with missing $Y$ later on. Let binary $R_i$ indicate whether the entire covariate vector $X_i$ is observed for patient $i$, where $i = 1, \hdots, n$. Let $X_i^{(mis)}$ and $X_i^{(obs)}$ correspond to the missing and observed entries in $X_i$ respectively. We will assume that observations are independent across $i$, although our results can be extended to settings with correlation across $i$. Additionally, we will assume that the data are missing at random (MAR) as defined in \citet{Little2002}, where missingness may depend only on fully-observed variables. We suppose our interest is in parameter $\theta$ corresponding to the assumed distribution for $Y \vert X$. \\
\indent Multiple imputation strategies attempt to draw multiple potential values for $X_i^{(mis)}$ from the posterior predictive distribution $f(X_i^{(mis)} \vert X_i^{(obs)}, Y_i)$ as follows:
\begin{align} \label{eqJoint}
& f(X_i^{(mis)} \vert X_i^{(obs)}, Y_i) \hspace{0.1cm} \propto \hspace{0.1cm} f(Y_i \vert X_i) f(X_i^{(mis)} \vert X_i^{(obs)})
\end{align}
Obtaining a draw from \ref{eqJoint} directly can be difficult, since the distribution is only known up to proportionality. Usual MICE imputation would attempt to approximate a draw from \ref{eqJoint} by drawing missing covariates from a series of simpler distributions. An alternative strategy for approximating a draw from \ref{eqJoint} is via importance sampling as discussed in \citet{Little2002}, where we first draw multiple times from $f(X_i^{(mis)} \vert X_i^{(obs)})$. Note that this distribution does not condition on $Y$. Then, we choose a \textit{single} imputation of $X_i^{(mis)}$ from these draws using a multinomial distribution where we select the $k^{th}$ draw with probability proportional to $f(Y_i \vert X_{ik})$ and where $X_{ik}$ corresponds to the $k^{th}$ draw of $X_i^{(mis)}$. Inference for either approach could then proceed by constructing \textit{multiple} imputed datasets, fitting the model of interest to each dataset, and combining inference across imputed datasets using Rubin's combining rules \citep{Little2002}. As shown in simulations, this approach can have good performance, but it can involve taking many, many draws from $f(X_i^{(mis)} \vert X_i^{(obs)})$, which can increase the computational burden. 

\subsection{Proposed imputation strategy} \label{proposed}
\begin{figure}[htbp!]
\caption{Diagram of Proposed Imputation Strategy*}
\centering
\includegraphics[trim={0cm 0cm 0cm 0cm}, clip, width=5in]{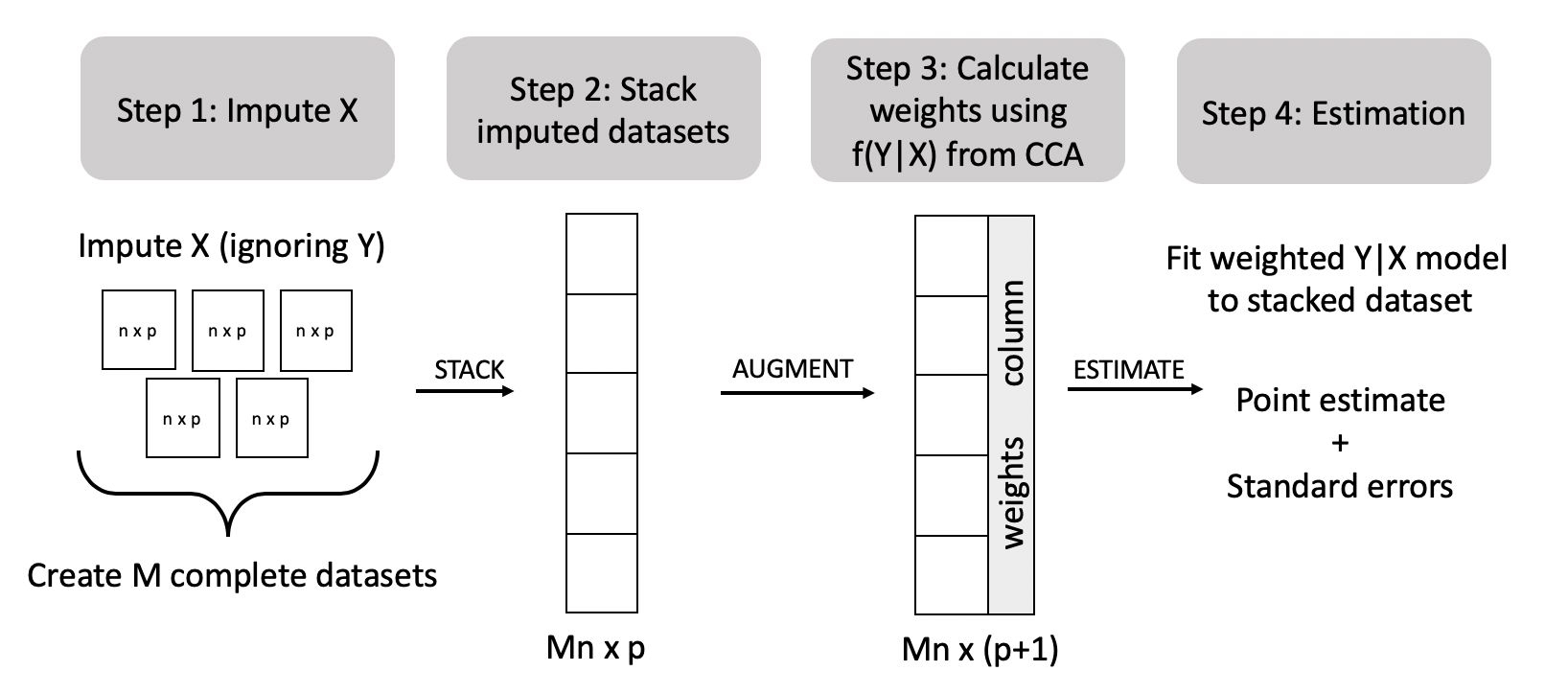}
\caption*{\footnotesize *CCA = complete case analysis}
\label{fig:diagram}
\end{figure}

\indent Rather than taking multiple draws from $f(X_i^{(mis)} \vert X_i^{(obs)})$ to obtain a \textit{single} imputation from \ref{eqJoint}, we propose using all those draws as our multiple imputations and weighting them proportional to $f(Y_i \vert X_i)$ in the final analysis, where weights are scaled to sum to 1 across imputations. Weights, therefore, are defined \textit{across} imputed datasets rather than \textit{within} imputed datasets. In order to make inference about $\theta$, we perform the following steps as shown in \textbf{Figure \ref{fig:diagram}}. We provide example R code for implementation in \textbf{Web Appendix 3}. 

\noindent $\bullet$ \underline{Step 1: Impute missingness in covariates ignoring $Y$}\\
\noindent In this step, we obtain the multiple imputations of $X_i$ from an assumed distribution for $f(X_i^{(mis)} \vert X_i^{(obs)})$, which in practice can be implemented using MICE by specifying regression models for each covariate with missingness given the other covariates but \textit{not including} the outcome. An additional complication arises when we also have missingness in $Y$. In this case, we can proceed as above to obtain imputations of $X$ ignoring $Y$ and then impute missing values of $Y$ from $f(Y \vert X)$ for each imputed dataset.  \\
\noindent $\bullet$  \underline{Step 2: Stack imputations}\\
\noindent We obtain a stacked version of the data, where each of the $M$ imputed datasets of size $n \times p$ are stacked on top of each other to form a $Mn \times p$ dataset, called the ``tall stack." An alternative stacking strategy is to include patients with fully-observed data only once in the stacked dataset. If $n_1$ is the number of patients with fully-observed data, this will result in a stacked dataset with $n_1 + (n-n_1)M$ rows, called the ``short stack." In settings where $n$ or $M$ is large, this may be a more memory- and computationally-efficient stacking strategy and should have no impact on resulting inference for appropriately defined weights.\\
\noindent $\bullet$  \underline{Step 3: Assign weights}\\
\noindent For usual data analysis of stacked multiple imputations, we augment the stacked dataset with weights defined for each row as 1 divided by the number of times that patient appears in the stacked dataset. For our modified imputation stacking approach, we augment the stacked dataset with a weight column, where weights are defined to be proportional to $f(Y_i \vert X_i)$. In practice, $f(Y_i \vert X_i) = \int f(Y_i \vert X_i; \theta) f(\theta \vert X_i^{(obs)}, Y_i) d\theta $ may be hard to calculate, since it involves integrating out the corresponding parameter. Instead, we replace $f(Y_i \vert X_i)$ with $f(Y_i \vert X_i, R_i =  1; \hat \theta_{cc})$ where $\hat \theta_{cc}$ is the estimated $\theta$ obtained from complete case analysis (CCA) for $Y \vert X$ (fit $Y \vert X$ to data from patients without any missingness). We define weights using complete case data following logic in \textbf{Section \ref{missingnessY}}. For the row corresponding to the $m^{th}$ imputation for the $i^{th}$ patient and corresponding imputed $X_{im}$, assign weight
\begin{align*}
& w_{im} = \frac{f(Y_i \vert X_{im}, R = 1; \hat \theta_{cc})}{\sum_{j=1}^M f(Y_i \vert X_{ij}, R = 1; \hat \theta_{cc})}
\end{align*}
For patients with fully-observed data, define $X_{im}$ to be equal to $X_i$. The resulting weights for patients with fully-observed data will be constant and equal to 1 divided by the number of times the patient appears in the stacked dataset (may be 1 or $M$). An alternative weighting strategy is to define weights as $w_{im} = \frac{f(Y_i \vert X_{im}, R = 1; \theta^m_{cc})}{\sum_{j=1}^M f(Y_i \vert X_{ij}, R = 1;  \theta^j_{cc})}$ where $\theta^j_{cc}$ is a draw of the complete-case $\theta$ rather than the MLE. In practice, there may be little difference between the two approaches, but the difference will likely be larger for smaller complete case samples. \\
\noindent $\bullet$  \underline{Step 4: Estimate $\theta$}\\
\noindent Estimate $\theta$ by fitting a weighted model for $Y \vert X$ to the stacked dataset with weights $w$. We describe how to estimate corresponding standard errors in next following section.

\subsection{Missingness dependent on Y} \label{missingnessY}
\indent Now, we consider the particular case where missingness is MAR dependent on $Y$. In this case, the proposed imputation strategy ignoring $Y$ induces a missing not at random (MNAR) mechanism when missingness is expressed only as a function of $X$ \citep{Little2002}. Therefore, additional thought is needed to assess whether it is appropriate to impute missing $X$ using the proposed approach when missingness depends explicitly on $Y$. We note that under MAR dependent on $Y$, 
\begin{align*}
&f(X^{(mis)} \vert X^{(obs)}, Y, R = 1)  = f(X^{(mis)} \vert X^{(obs)}, Y) \\
\text{ but } &f(Y \vert X, R = 1) \neq  f(Y \vert X) \text{ and } f(X^{(mis)} \vert X^{(obs)}, R = 1) \neq f(X^{(mis)} \vert X^{(obs)})
\end{align*}
Complete case analysis will produce biased results for the parameter of $Y \vert X$, and it may also produce biased results for the parameter related to $f(X^{(mis)} \vert X^{(obs)}, R = 1)$. However, we note that 
\begin{align*}
&f(X^{(mis)} \vert X^{(obs)}, Y) = f(X^{(mis)} \vert X^{(obs)}, Y, R = 1) \propto  f(Y \vert X, R = 1) f(X^{(mis)} \vert X^{(obs)}, R = 1)\
\end{align*}
This suggests that we might impute $X^{(mis)}$ by drawing $X^{(mis)}$ from $f(X^{(mis)} \vert X^{(obs)}, R = 1)$ and then weighting by $f(Y \vert X, R = 1)$ to produce valid results even if missingness depends on $Y$. Interestingly, we can use the proposed methods to obtain imputations under MAR dependent on $Y$ even though parameter estimates/draws in the imputation and weighting steps \textit{individually} are expected to be biased. Roughly, we can think of the biases as ``cancelling each other out." In practice, MICE does not exactly impute each missing $X^p$ using drawn parameters conditioning on $R = 1$ (overall complete case data) as suggested by the above equation. Instead, the algorithm draws parameters for imputation of a given covariate $X^p$ using data from patients with $X^p$ fully observed. While this results in a potential for residual bias in estimating outcome model parameters downstream, we expect this bias to be generally small as demonstrated in our simulations.

\section{Estimating Standard Errors} \label{stderrs}
A major drawback of the stacked imputation approach in general is the difficulty in estimating standard errors. Conventional estimators such as sandwich estimators only account for the so-called ``within-imputation" variation, ignoring the ``between-imputation" variation \citep{Wood2008}. \citet{Wood2008} proposed a strategy for scaling up the standard errors obtained from fitting a model to the stacked data. Standard errors associated with covariate $X^p$ are obtained by fitting a model for $Y \vert X$ and weighting each row of the stacked data by $\frac{1-f_p}{M}$, where $f_p$ is the fraction of missing information in $X^p$. The fraction of missing information $f_p$ is roughly estimated as the proportion of patients with missing values for $X^p$. This strategy requires the model of interest to be re-fit multiple times to obtain standard errors for each $X^p$. Alternatively, we can obtain similar standard errors by post-multiplying the variance associated with covariate $X^p$ by $\frac{M}{1-f_p}$ after fitting a single regression model weighted by $1/M$. This approach from \citet{Wood2008} is motivated under MCAR missingness and simple to implement, but its ability to estimate standard errors in other missingness settings is unclear. \\
\indent \citet{Yang2016} and \citet{Kim2011} developed a stacked imputation strategy in the survey sampling context called fractional multiple imputation. Estimation proceeds using an iterative algorithm in which we define weights as a function of the analysis/imputation methods and survey weights, estimate parameters of interest, re-estimate weights, etc. Standard errors are then estimated using a jackknife-type approach. This estimator can be complicated and computationally expensive to estimate, and the lack of available software for general parametric fractional imputation severely limits its ability to be used in practice. \\
\indent Another strategy in the literature for estimating standard errors for stacked multiple imputation was developed in \citet{Robins2000} and more recently applied in \citet{Hughes2016}. This estimator requires score and information matrices for \textit{both} the imputation and analysis models. Additionally, the estimator itself can be complicated to conceptualize and compute, and no standard software exists to make such calculations routine. This approach also requires that the imputation models are standard parametric models from which we can obtain score and information matrices, which excludes many popular non-parametric imputation strategies such as random forests or predictive mean matching. Given the complexity that serves as a barrier to general use of this estimator, we chose not to implement the methods in \citet{Robins2000} and \citet{Kim2011} in our simulations later on. \\
\indent We propose an alternative strategy for estimating standard errors that, like the method in \citet{Robins2000}, involves the score and information matrices from the outcome model. Unlike \citet{Robins2000}, however, we \textit{do not} require information about the imputation distributions. Our proposed estimator can be applied in usual imputation stacking settings and in our modified imputation stacking approach that explicitly incorporates the outcome model into defining the weights. Like standard errors from Rubin's rules (but unlike \citet{Robins2000}), our estimator is not guaranteed to have good performance when imputation and analysis models are uncongenial. Our proposed estimator takes advantage of the complete information principle discussed in \citet{Louis1982}, namely $I_{obs}(\theta) = I_{com}(\theta)  - I_{mis}(\theta)$, where $I_{obs}$ is the observed data information matrix (the target), $I_{com}$ is the expected complete data information matrix given the observed data, and $I_{mis}$ is the expected missing information given the observed data. While $I_{obs}$ can be difficult to estimate directly, $I_{com}$ and $I_{mis}$ may be more readily estimated. First, we will assume data are independent across values of $i$. Let $J^i_{com}$ correspond to the complete data Fisher information matrix contribution for patient $i$, and let $U^i_{com}$ be the corresponding score matrix contribution for patient $i$. See \textbf{Web Appendix 2} for an example. \citet{Wei1990} proposed a Monte Carlo version of the estimator developed in \citet{Louis1982} that involves averaging the estimated $I_{com}$ and $I_{mis}$ across multiple imputations of the data. Using a similar strategy, we propose a generalization of the estimator in \citet{Louis1982} that allows for individual and imputation-specific weights, $w_{im}$, and involves averaging across multiple imputations. With imputation as in \textbf{Figure \ref{fig:diagram}}, $w_{im}$ corresponds to the augmented weight in Step 3. With general multiple imputation, we can define $w_{im}$ for each $i$ as the number of times that subject appears in the stacked dataset ($M$ for tall stack, 1 for short stack). Let $X_{im}$ denote the $m^{th}$ imputation of $X_i$. For subjects with fully-observed $X_i$, define $X_{im} = X_i$. As shown in \textbf{Web Appendix 1}, we can express
\begin{align} \label{louis}
& I_{obs}(\hat \theta) \approx \sum_i E_{\hat \theta} \left[ J^i_{com}(X_i,Y_i) \vert X_i^{obs},Y_i \right]  -  \sum_i  Var_{\hat \theta} \left[ U^i_{com}(X_i,Y_i)\vert X_i^{obs},Y_i \ \right] \nonumber\\
& \approx \sum_i \sum_m w_{im} J^i_{com}(X_{im}, Y_i)  -  \sum_i  \sum_m w_{im} \left[ U^i_{com}(X_{im}, Y_i) - \bar U^k_{com}  \right]^{\otimes 2} 
\end{align}
where $\bar U^k_{com}   = \sum_j w_{kj} U^k_{com}(X_{kj}, Y_k)$ and where $\hat \theta$ is the point estimate obtained from fitting the weighted model for $Y \vert X$ on the stacked data. The first element in the above equation is the weighted complete data information matrix for the outcome model evaluated using the stacked dataset. The second term is the weighted variance of $U_{com}^i$ summed over the patients $i$ with imputed data. Given the equations for the complete data score and information matrix for an individual under the outcome model, these quantities can be easily calculated using the stacked data. We have developed an accompanying R package \textit{StackImpute} that provides functions for calculating these standard errors for several common regression models including generalized linear models and Cox proportional hazards models. 

\section{Simulations} \label{sims}
In this section, we provide results from a simulation study exploring the performance of the proposed imputation strategy and corresponding standard error estimator in terms of bias, coverage, and empirical variances of point estimates. This simulation study is broken up into four scenarios: (1) Gaussian $Y$ with missingness in a single covariate, (2) binary $Y$ with missingness in two covariates,  (3) Gaussian $Y$ with missingness in a single covariate and interactions in the outcome model, and (4) censored survival-type $Y$ with missingness in a single covariate. We consider four different missingness mechanisms: MCAR, MAR dependent on $X$, MAR dependent on $Y$, and MAR dependent on both $X$ and $Y$.

\subsection{Simulation set-up}
In all four scenarios, we generated 500 simulated datasets of 2000 patients each. Simulations then proceeded as follows:\\
\\
\textbf{Scenario 1:} Gaussian $Y \vert X_1, X_2$ with missingness in $X_2$\\
We generate covariates $X_1$ and $X_2$ from a multivariate normal distribution with mean 0, Var($X_1$) = 0.49, Var($X_2$) = 0.09, and covariance of 0.12. We then generated $Y \vert X_1, X_2 \sim N(0.53 X_1 + 1.25 X_2, 0.55)$. Roughly 50\% missingness was generated in $X_2$ under the model $\text{logit}(P(\text{$X_2$ observed} \vert X_1, Y)) = \phi_0 + \phi_1 X_1 + \phi_2 Y$ with values $\phi = \{ (0,0,0), (0,1,0), (0,0,1), (0,1,-1)\}$. \\
\\
\textbf{Scenario 2:} Binary $Y \vert X_1, X_2, X_3$ with missingness in $X_2, X_3$\\
We generate covariates $X_1$, $X_2$, and $X_3$ from a multivariate normal distribution with mean 0, unit variances, and pairwise covariance of 0.3.  We then generated binary $Y$ using the relation $\text{logit}(P(Y=1 \vert X_1, X_2, X_3)) = 0.5 + 0.5 X_1 + 0.5 X_2 + 0.5 X_3$. Missingness in $X_2$ was generated using the model from Scenario 1 with $\phi = \{ (0.5,0,0), (0.5,1,0), (0.5,0,1), (0.5,1,-1)\}$, and missingness is independent of $X_3$. We then induced 30\% MCAR missingness for $X_3$. This resulted in roughly 40\% of patients having complete data.\\
\\
\textbf{Scenario 3:} Gaussian $Y \vert X_1, X_2, X_1\times X_2$ with missingness in $X_2$\\
We generate covariates $X_1$ and $X_2$ from a multivariate normal distribution with mean 0, Var($X_1$) = 0.81, Var($X_2$) = 1.21, and covariance of 0.59. We then generated $Y \vert X_1, X_2 \sim N(0 + X_1 +  X_2 + X_1 \times X_2, 1)$. We generate missingness in $X_2$ as in Scenario 1. \\
\\
\textbf{Scenario 4:} Exponential $T \vert X_1, X_2$ with missingness in $X_2$ and uniform censoring\\
We generate covariates $X_1$ and $X_2$ from a multivariate normal distribution with mean 0, Var($X_1$) = 1, Var($X_2$) = 1, and covariance of 0.5. We then generated $T \vert X_1, X_2 $ to have an exponential distribution with scale parameter $e^{0.5 X_1 + 0.5 X_2}$. Uniform(0.2, 3) censoring was then imposed on $T$. Missingness in $X_2$ was generated using the same model as Scenario 1, with $\phi = \{ (0.5,0,0), (0.5,1,0)\}$. We did not explore missingness related to the outcome. \\
\\
\indent Once the data were simulated, we performed multiple imputation of the missing values of $X$ using the methods described in this paper to obtain $M=50$ multiple imputations. We then analyzed the results fitting the \textit{correct} outcome model either using Rubin's combining rules or the proposed stacking method with standard errors estimated using various strategies including the standard sandwich estimator from the R package \textit{sandwich}, the method in \citet{Wood2008}, and our estimator in \ref{louis}. In Scenario 4, stacked analysis weights were defined based on a Cox model fit to the complete case data. From this fit, we obtained the Breslow estimator for the cumulative baseline hazard and defined a piecewise constant baseline hazard that integrated to produce the estimated cumulative baseline hazard. Weights proportional to $f(Y \vert X)$ could then be calculated.

\subsection{Simulation results}
\indent \textbf{Table \ref{biastable}} shows the estimated bias of outcome model parameters across 500 simulated datasets. Complete case analysis shows substantial bias in Scenarios 1 and 3 whenever missingness depends on $Y$. In Scenario 2, complete case analysis is biased only when missingness depends on both $Y$ and covariate values, following well-known properties of logistic regression under case-control sampling. MICE with $Y$ in the imputation model resulted in correctly-specified imputation models in Scenario 1 only. Evidence of resulting bias can be see for Scenarios 3 and 4. For these settings, imputation using SMC-FCS as in \citet{Bartlett2014} tends to produce little bias since imputation was performed using the ``correct"  distributions. Stacked and $1/M$ weighted analysis using MICE imputations conditioning on $Y$ produced very similar results in terms of bias to analysis of these imputations using Rubin's rules. Analysis based on stacking imputations obtained without $Y$ and weighting rows by $f(Y \vert X)$ produced little bias across simulation scenarios. \\
\indent \textbf{Table \ref{empvartable}} shows the relative empirical variance of point estimates (compared to analysis of the full data) across 500 simulated datasets. Empirical variances for the stacking methods with $1/M$ weighting tend to be consistent with standard errors estimated using Rubin's rules. Stacking of imputations ignoring $X$ and then weighting by $f(Y \vert X)$ produces similar empirical variances to the SMC-FCS method from \citet{Bartlett2014}. Empirical variances for SMC-FCS can have higher or lower empirical standard errors relative to MICE methods that use regression model approximations of the conditional distributions for imputation.   \\
\indent \textbf{Table \ref{varestimationtable}} shows the average estimated standard errors and the 95\% confidence interval coverage rates for different variance estimation strategies based on stacked data analysis. The sandwich estimator applied to the stacked and weighted data tends to strongly under-estimate variance. This is because this estimator accounts for ``within-imputation" variation but does not appropriately address ``between-imputation" variation. The method in \citet{Wood2008} is an improvement over the sandwich estimator, but this estimator can result in sub-optimal coverage even in the MCAR setting. The \citet{Wood2008} method produced overly-conservative standard errors for imputed covariates. The proposed estimation strategy in \ref{louis} produced nominal coverage and standard error estimates near those obtained using the method in \citet{Bartlett2014}, here viewed as a gold standard. In the proposed algorithm, weights were obtained using parameter \textit{estimates} from a complete case fit for $f(Y \vert X)$ rather than parameter \textit{draws}. Drawing the corresponding parameter when defining weights produced very similar results to using estimates in this simulation.

 \begin{table}[htb!]
\small
\caption{ Bias of outcome model parameters under various imputation strategies and outcome model settings. Results across 500 simulations are presented. Biases greater than 0.05 are shaded. In all settings, $X_1$ was fully-observed and $X_2$ and possibly $X_3$ were imputed. All biases were multiplied by 100. }
\begin{tabular}{lcccc|cccc}
\hline
\\
 &  \multicolumn{4}{c}{Bias $\times 100$ in effect of $X_1$}&  \multicolumn{4}{c}{Bias $\times 100$ in effect of $X_2$}  \\
 \\
 \hline
Missingness:   & MCAR & $X_1$ & $Y$ & $X_1,Y$  & MCAR & $X_1$ & $Y$ & $X_1,Y$\\
\hline
\\
 & \multicolumn{8}{c}{Scenario 1: Linear Regression} \\
 Full Data & 0.02 & 0.01 & 0.14 & 0.28 & -0.05 & -0.15 & -0.17 & -0.20 \\
Complete Case & -0.03 & -0.05 &  \cellcolor{light-gray}-5.18 &  \cellcolor{light-gray} 5.29 & -0.16 &  0.18 & \cellcolor{light-gray} -13.11 &  \cellcolor{light-gray}-13.59  \\
MICE with $Y$* & \\
 \hspace{0.2cm} $\drsh$ Rubin's rules & 0.08 &  0.03 & 0.28 & 0.36 &  -0.41 & 0.02 & -0.75 &  -0.30\\
 \hspace{0.2cm} $\drsh$ Stacked, 1/M weighted & 0.11& 0.07 & 0.32 & 0.39 &  -0.53 & -0.12 &  -0.88 & -0.41  \\
MICE without $Y$* & \\
 \hspace{0.2cm} $\drsh$ Rubin's rules & \cellcolor{light-gray}16.1&   \cellcolor{light-gray}16.1 &  \cellcolor{light-gray}18.48 &  \cellcolor{light-gray}18.0 &   \cellcolor{light-gray} -62.6 & \cellcolor{light-gray} -62.3 &  \cellcolor{light-gray}-69.09 &  \cellcolor{light-gray}-69.4 \\
 \hspace{0.2cm} $\drsh$ Stacked, $f(Y \vert X)$ weighted &  0.32  & 0.27 & 0.60 & 0.66 & -1.36 & -0.88 & -1.85 & -1.46 \\
MICE multinomial**  & 0.27 & 0.23 &  0.54 & 0.62 & -1.19 & -0.73 &  -1.67 &-1.31 \\
\\
 & \multicolumn{8}{c}{Scenario 2: Logistic Regression} \\
 Full Data & 0.34 &  -0.03 & 0.09 & 0.13 & 0.24 &-0.09 & 0.22 &  0.12 \\
 Complete Case &0.75 &  0.37 & -0.12 &  \cellcolor{light-gray} 21.0 & 0.18 & -0.09 & 0.56 & 0.32  \\
MICE with $Y$ &\\
 \hspace{0.2cm} $\drsh$ Rubin's rules &  0.35 & -0.08 &  0.05 & -0.07 &  -0.17 & -0.60 & 0.17 & -0.53  \\
 \hspace{0.2cm} $\drsh$ Stacked, 1/M weighted &  0.35 &  -0.08 & 0.04 & -0.09 &  -0.26 &  -0.73 & 0.10 & -0.72\\
MICE without $Y$  &\\
 \hspace{0.2cm} $\drsh$ Rubin's rules &   \cellcolor{light-gray}5.85 &   \cellcolor{light-gray}5.87 &  \cellcolor{light-gray}5.01 &  \cellcolor{light-gray}6.49 &  \cellcolor{light-gray}-18.49 &  \cellcolor{light-gray} -20.8 & \cellcolor{light-gray}-14.5 & \cellcolor{light-gray} -26.6  \\
 \hspace{0.2cm} $\drsh$ Stacked, $f(Y \vert X)$ weighted  & 0.49 & 0.11 & 0.13 & 0.30 &  -0.25 & -0.61 & 0.12 & -0.43 \\
Bartlett et al. (2014) $\bowtie$& 0.42 & 0.05 & 0.09 & 0.08 &  0.12 &-0.31 &0.30 & -0.19    \\
\\
 & \multicolumn{8}{c}{Scenario 3: Linear Regression with Interaction} \\
  Full Data & 0.10 &0.10 & 0.29 &-0.22 &-0.14 & -0.04 &  -0.30 & 0.26  \\
 Complete Case & 0.21 & -0.10 &  \cellcolor{light-gray}-8.97 &  -0.58 & -0.36 &  -0.09 &  \cellcolor{light-gray} -9.90 & \cellcolor{light-gray}-14.88   \\
MICE with $Y$ & \\
 \hspace{0.2cm} $\drsh$ Rubin's rules & -2.12 &  \cellcolor{light-gray}-13.9 & -4.73 & \cellcolor{light-gray} -7.99 &  \cellcolor{light-gray} -12.28 &  \cellcolor{light-gray}13.14 &  -1.35 & -3.97   \\
 \hspace{0.2cm} $\drsh$ Stacked, 1/M weighted & -2.07 & \cellcolor{light-gray} -13.95 & -4.70 & \cellcolor{light-gray} -7.82 & \cellcolor{light-gray}   \cellcolor{light-gray}-12.40 & \cellcolor{light-gray}  13.11 & -1.38 &  -4.29 \\
MICE with $Y$ + interactions &  -2.75 &  \cellcolor{light-gray}18.93 &  \cellcolor{light-gray}-10.05 &  \cellcolor{light-gray}-17.52 &  \cellcolor{light-gray}-10.28 &  \cellcolor{light-gray}21.35 &  \cellcolor{light-gray}5.93 & \cellcolor{light-gray} -10.14   \\
MICE without $Y$  &  \\\
 \hspace{0.2cm} $\drsh$ Rubin's rules & \cellcolor{light-gray}36.8 & \cellcolor{light-gray} 24.13 & \cellcolor{light-gray}  16.84 & \cellcolor{light-gray}81.70 & \cellcolor{light-gray} -50.20 &  \cellcolor{light-gray} -32.75 &  \cellcolor{light-gray}-35.32 & \cellcolor{light-gray} -70.16   \\
 \hspace{0.2cm} $\drsh$ Stacked, $f(Y \vert X)$ weighted & 0.05 & 0.05 & -1.22 & -1.24 &  -0.10 & -0.08 & - 1.37 & 0.01 \\
Bartlett et al. (2014) &  0.38 & 0.19 & 0.35 & 0.40 & -0.49 & -0.22 & -0.50 & 0.16   \\
\\
 & \multicolumn{8}{c}{Scenario 4: Cox Proportional Hazards Regression} \\
  Full Data &  0.12  &  0.04 &  - & - &   0.18& 0.10 & - & -  \\
 Complete Case & 0.12  &   0.07&  - & - & 0.07  &  0.26 & - & -  \\
MICE with $Y$  & \\
 \hspace{0.2cm} $\drsh$ Rubin's rules &  -1.62  & -1.65  &  - & - &   -4.18& 0.37  & - & -  \\
 \hspace{0.2cm} $\drsh$ Stacked, 1/M weighted & -1.61  &  -1.59 &  - & - &  -4.30&  0.27 & - & - \\
MICE without $Y$  &  \\
 \hspace{0.2cm} $\drsh$ Rubin's rules & 0.48 &   1.58&  - & - & \cellcolor{light-gray} -27.2  & \cellcolor{light-gray}  -25.02 & - & -  \\
 \hspace{0.2cm} $\drsh$ Stacked, $f(Y \vert X)$ weighted  &  0.15 &  0.56 &  - & - &  -0.30 &  -2.43 & - & -  \\
Bartlett et al. (2014)  &  0.15 & -0.05 & - & - & 0.03 & 0.25 & - & - \\
\\
\hline
\end{tabular} \label{biastable}
\caption*{ \footnotesize *MICE either including or excluding $Y$ from the linear regression imputation models. An interaction between $Y$ and $X_1$ was included in some settings for Scenario 3. MICE with $Y$ for Scenario 4 followed recommendations in \citet{White2009}. Unless otherwise specified, MICE imputations were analyzed using Rubin's rules.  \\ 
** Impute many times ignoring $Y$ and choose imputation $k$ with probability proportional to $f(Y\vert X)$. \\
 $\bowtie$ $X^p$ imputed from distribution $ \propto f(Y \vert X) f(X^p \vert X^{-p})$ using R package \textit{smcfcs}. Then, apply Rubin's rules.\\
}
\end{table}

 \begin{table}[htb!]
\small
\caption{ Relative empirical variance of outcome model parameters under various imputation strategies and outcome model settings (relative to full data without missingness). Results across 500 simulations are presented. In all settings, $X_1$ was fully-observed and $X_2$ and possibly $X_3$ were imputed.}
\begin{tabular}{lcccc|cccc}
\hline
\\
 &  \multicolumn{4}{c}{Relative variance for effect of $X_1$}&  \multicolumn{4}{c}{Relative variance for effect of $X_2$}  \\
 \\
 \hline
Missingness:   & MCAR & $X_1$ & $Y$ & $X_1,Y$  & MCAR & $X_1$ & $Y$ & $X_1,Y$\\
\hline
\\
 & \multicolumn{8}{c}{Scenario 1: Linear Regression} \\
 Full Data & 1.00 & 1.00 & 1.00 & 1.00 & 1.00 & 1.00 & 1.00 & 1.00 \\
Complete Case & 2.06 & 2.07 & 1.87 & 1.85 & 1.88 &   2.09 & 1.75 &1.73  \\
MICE with $Y$* & \\
 \hspace{0.2cm} $\drsh$ Rubin's rules & 1.35 &  1.37 & 1.45 &  1.31 &  1.70 & 1.85 & 1.98 &  1.90\\
 \hspace{0.2cm} $\drsh$ Stacked, 1/M weighted & 1.35 &   1.37 & 1.45 & 1.31 & 1.70  & 1.85 &  1.97 & 1.90 \\
Stacked, $f(Y \vert X)$ weighted $\dagger$ &  1.34 &1.37 &1.45 & 1.31 &  1.69 & 1.83  & 1.95 & 1.89\\
\\
 & \multicolumn{8}{c}{Scenario 2: Logistic Regression} \\
 Full Data & 1.00 & 1.00 & 1.00 & 1.00 & 1.00 & 1.00 & 1.00 & 1.00 \\
Complete Case & 2.52 & 2.29 &2.02 & 4.08 &  2.36 &  2.46 & 2.15 & 3.66 \\
MICE with $Y$ &\\
 \hspace{0.2cm} $\drsh$ Rubin's rules &  1.08 &1.08 & 1.04 & 1.13 & 1.64 &1.64 & 1.45 &  2.35   \\
 \hspace{0.2cm} $\drsh$ Stacked, 1/M weighted  & 1.08 & 1.07 & 1.04 & 1.12 &  1.63 &  1.63 &  1.45 & 2.33 \\
Stacked, $f(Y \vert X)$ weighted  & 1.09 & 1.08 & 1.03 &1.14 & 1.78 & 1.82 & 1.55 & 2.77   \\
Bartlett et al. (2014) $\bowtie$& 1.09 &1.09 &  1.05 & 1.14 &1.73 &1.74 & 1.52 & 2.58   \\
\\
 & \multicolumn{8}{c}{Scenario 3: Linear Regression with Interaction} \\
 Full Data & 1.00 & 1.00 & 1.00 & 1.00 & 1.00 & 1.00 & 1.00 & 1.00 \\
 Complete Case & 2.14&   2.13 & 1.78 & 2.37 & 2.11 & 2.04 & 1.83 &2.50  \\
MICE with $Y$ &\\
 \hspace{0.2cm} $\drsh$ Rubin's rules & 2.85 & 2.12 &  1.34 & 5.20 & 3.16 & 3.35 & 1.62 &  4.02\\
 \hspace{0.2cm} $\drsh$ Stacked, 1/M weighted & 2.85 &  2.12 & 1.34 & 5.21 & 3.16 &3.35 & 1.62 &4.05  \\
Stacked, $f(Y \vert X)$ weighted  &  1.50 & 1.40 & 1.26 &2.07 &  1.74 & 1.71 & 1.60 & 2.06 \\
Bartlett et al. (2014) & 1.52 &  1.46 & 1.29 &2.07 &  1.75 & 1.60 &  1.55 & 1.99 \\
\\
 & \multicolumn{8}{c}{Scenario 4: Cox Proportional Hazards Regression} \\
  Full Data &  1.00 & 1.00&  - & - & 1.00 & 1.00 & - & -  \\
 Complete Case &  1.85 & 2.20  &  - & - &   2.13& 1.70  & - & -  \\
MICE with $Y$ & \\
 \hspace{0.2cm} $\drsh$ Rubin's rules &  1.06  &  1.13 &  - & - & 1.62  &  1.57 & - & -  \\
 \hspace{0.2cm} $\drsh$ Stacked, 1/M weighted &   \\
Stacked, $f(Y \vert X)$ weighted  & 1.13 &  1.13 &  - & - &  1.91 &  1.54 & - & -  \\
Bartlett et al. (2014)  &  1.15 & 1.19 & - & - & 2.02 & 1.76 & - & -\\
\\
\hline
\end{tabular} \label{empvartable}
\caption*{ \footnotesize *MICE including $Y$ in the linear regression imputation models. MICE with $Y$ for Scenario 4 followed recommendations in \citet{White2009}.\\ 
 $\dagger$ Stacked version of MICE without $Y$ and using weights proportional $f(Y\vert X)$. \\
 $\bowtie$ $X^p$ imputed from distribution $\propto f(Y \vert X) f(X^p \vert X^{-p})$ using R package \textit{smcfcs}. Then, apply Rubin's rules.\\
}
\end{table}

 \begin{table}[htb!]
\small
\caption{ Average estimated variance (coverage of 95\% confidence intervals) for Scenario 2 outcome model parameters under various imputation strategies. Results across 500 simulations are presented, and all elements in table have been multiplied by 100. $X_1$ was fully-observed and $X_2$, $X_3$ were imputed.}
\begin{tabular}{lcccc|cccc}
\hline
\\
 &  \multicolumn{4}{c}{Variance (coverage) for effect of $X_1$}&  \multicolumn{4}{c}{Variance (coverage) for effect of $X_2$}  \\
 \\
 \hline
Missingness:   & MCAR & $X_1$ & $Y$ & $X_1,Y$  & MCAR & $X_1$ & $Y$ & $X_1,Y$\\
 \hline
\\
 & \multicolumn{8}{c}{MICE with $Y$} \\
 Empirical & 0.35 (---) & 0.36 (---) & 0.34 (---)& 0.36 (---)& 0.54 (---)& 0.58(---)& 0.48 (---)& 0.72 (---)\\
Rubin's rules & 0.33 (96) & 0.39 (94) & 0.31 (96) & 0.34 (96) & 0.50 (95) & 0.57 (95) &0.52 (95) & 0.72 (96) \\
\\
 & \multicolumn{8}{c}{MICE with $Y$, Stacked and 1/$M$ weighted } \\
Empirical & 0.33 (---) & 0.39 (---) & 0.31 (---)& 0.34(---)& 0.51 (---)& 0.56(---)& 0.52 (---)& 0.72 (---)\\
Sandwich** & 0.01 (25) & 0.01 (21) & 0.01 (24) & 0.01 (21) & 0.01 (16)& 0.01 (16) & 0.01 (18) &  0.01 (13)    \\
Wood et al. (2008)  & 0.33 (95) & 0.33 (93) & 0.33 (96) & 0.33 (94) & 0.87 (99)&0.82 (98) & 1.26 (99) & 0.62 (94)     \\
Proposed \ref{louis}  & 0.34 (95) & 0.35 (93) & 0.34 (96) & 0.35 (95) & 0.53 (94) & 0.57 (96) & 0.47 (94) & 0.67 (95) \\
\\
 & \multicolumn{8}{c}{Bartlett et al. (2014)} \\
 Empirical & 0.33 (---) & 0.40 (---) & 0.31 (---)& 0.34 (---)& 0.53 (---)& 0.60(---)& 0.55 (---)& 0.79 (---)\\
Rubin's rules & 0.35 (95) & 0.36 (93) & 0.35 (96) & 0.36 (95) & 0.56 (95) & 0.60 (95) &0.50 (94) & 0.74 (93) \\
\\
 & \multicolumn{8}{c}{MICE without $Y$, Stacked and $f(Y \vert X)$ weighted } \\
Empirical & 0.33 (---) & 0.40 (---) & 0.31 (---)& 0.34 (---)& 0.54 (---)& 0.62 (---)& 0.56 (---)& 0.84 (---)\\
Sandwich & 0.01 (25)& 0.01 (18) & 0.01 (26) & 0.01 (20) &  0.01 (16) &0.01 (14) & 0.01 (15) & 0.01 (12)  \\
Wood et al. (2008) & 0.34 (95) & 0.33 (93) & 0.34 (96) & 0.34 (94) &  0.85 (98) & 0.80 (98) & 1.25 (99) & 0.60 (90) \\
Proposed \ref{louis} & 0.34 (94) & 0.35 (94) & 0.34 (96) & 0.35 (94) &  0.53 (94) &0.57 (95) & 0.47 (93) & 0.67 (92)  \\
 \hspace{0.2cm} $\drsh$ Draw $\theta$*&  0.34 (95) & 0.35 (93) & 0.34 (96) &  0.35 (95) & 0.52 (95) & 0.57 (94) & 0.47 (93) & 0.66 (92) \\
\\
\hline
\end{tabular} \label{varestimationtable}
\caption*{\footnotesize 
* Weights estimated using draws (one draw for each imputed dataset) of the complete case $\theta$. \\
** Standard errors estimated accounting for correlation between imputed datasets using the sandwich estimator implemented by R package \textit{sandwich}. 
}
\end{table}

\newpage
\section{Illustrative example: head and neck cancer survival}
\begin{figure}[htbp!]
\caption{HPV log-hazard ratio from Cox modeling of overall survival and time to recurrence using imputed head and neck cancer data }
\centering
        \includegraphics[trim={0cm 0.6cm 0cm 1cm}, clip, width=3in]{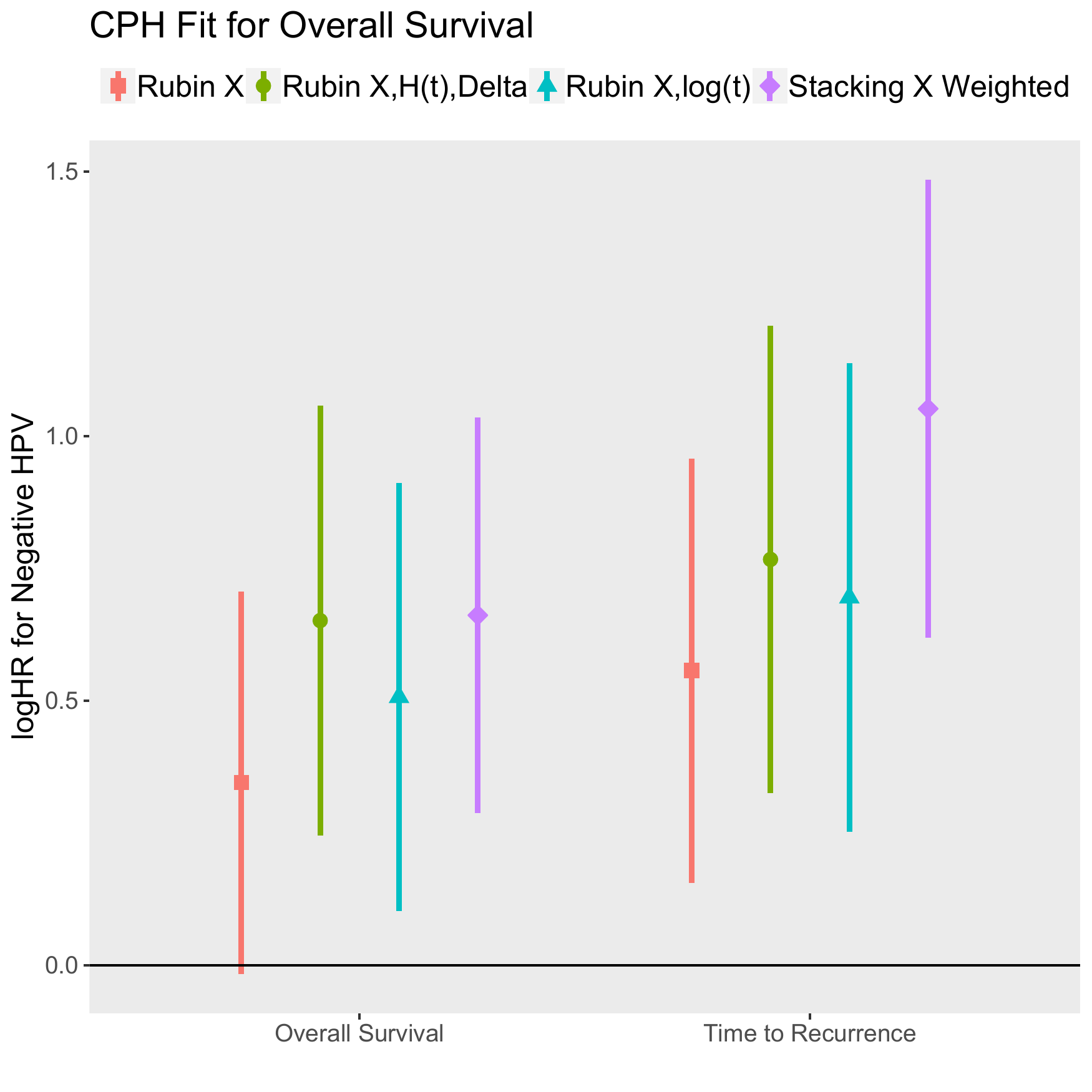}
        \label{hnscc}
\end{figure}

In this section, we illustrate the proposed methods for handling covariate missingness when we have a time-to-event outcome. In particular, we consider data from a study of 1226 patients treated for head and neck cancer at The University of Michigan. After initial treatment, consenting patients were followed for cancer recurrence and death. Smoking status (none, former, never), ACE27 comorbidities (none, mild, moderate, severe), HPV (human papillomavirus) status (positive, negative), age, cancer site (hypopharynx, larynx, oral cavity, oropharynx), and T stage (T0, T1, T2, T3) were recorded at baseline for the majority of patients, but T stage and HPV status were missing for roughly 30\% and 45\% of patients respectively. Small amounts of missingness were also present in smoking status and comorbidities. Additional study details can be found in \citet{Duffy2008} and \citet{Peterson2016}. \\
\indent We explore the impact of different imputation strategies on Cox proportional hazards model fits for overall survival and time to cancer recurrence. We note that a Cox proportional hazards mixture cure model would be more appropriate for time to cancer recurrence for head and neck cancer, but we will explore a standard Cox model fit for simplicity \citep{Beesley2016}. For each outcome model, our observed outcome can be written as $Y = (T, \delta)$, where $T$ is the event or censoring time for a given outcome event, and $\delta$ is the corresponding event/censoring indicator. We are interested in imputing missing values in $X$ (particularly, HPV status and T stage) using chained equations and somehow incorporating information in $Y$. \\
\indent Several methods exist in the literature for imputing missing covariates with time-to-event outcomes. \citet{VanBuuren1999} suggests imputing missing values in $X^p$ using a regression model with $X^{-p}$ and $\text{log}(T)$ as predictors, where $X^{-p}$ represents the covariates in $X$ excluding $X^p$. \citet{White2009} proposes imputation using predictors $X^{-p}$, $\delta$, and $H_0(T)$ as predictors, where $H_0(T)$ is an estimate of the cumulative baseline hazard for the event of interest. In practice,  \citet{White2009} suggests using the Nelson-Aalen estimate of the marginal cumulative hazard for imputation. We compared these imputation strategies to MICE imputation that entirely ignores the outcome variables $T$ and $\delta$. Imputation of HPV status assumed a logistic regression model structure, and imputation of all other variables assumed a multinomial regression. We then fit the outcome model of interest to each of the imputed datasets and obtained a single set of parameter estimates and standard errors using Rubin's combining rules \citep{Little2002}. \\
\indent Using imputations that were generated ignoring $Y = (T, \delta)$, we applied our proposed stacking and weighting strategy in \textbf{Figure \ref{fig:diagram}}, where we weighted each row proportional to $f(T_i, \delta_i \vert X_i) = \left[ \lambda_0(T_i) e^{\theta X_i} \right]^{\delta_i} e^{-\Lambda_0(T_i) e^{\theta X_i}}$ where $\lambda_0(t)$ and $\Lambda_0(t)$ are the baseline and cumulative baseline hazard functions respectively. These were obtained by fitting a Cox proportional hazards model to the imputed data to the complete case data. From there, we obtained the Breslow estimator for $\Lambda_0(t)$ and defined $\lambda_0(t)$ to be piecewise constant so that it integrated to $\Lambda_0(t)$. Standard errors for all stacked analyses were estimated using the  method in \ref{louis}. \\
\indent \textbf{Figure \ref{hnscc}} presents the resulting estimated HPV status log-odds ratio from Cox regressions for overall survival and time to recurrence outcomes adjusting for other patient-related factors. In both cases, imputation was performed using the overall survival outcome, so we might treat the time-to-recurrence analysis as a secondary analysis applied to previously imputed data, where the imputation and analyses models are not congenial. For the overall survival outcome, the proposed methods produced HPV status confidence intervals very near those obtained using Rubin's rules and MICE imputation as in \citet{White2009}. However, the stacked imputation method produces a much larger hazard ratio estimate for the time to recurrence outcome compared to all other methods. This difference may be because, unlike the other methods, our proposed method incorporates the assumed time-to-recurrence model structure into the imputation and, therefore, does not suffer from uncongeniality.

\section{Discussion}

\indent Multiple imputation using chained equations (MICE) is a popular and attractive approach for handling missing data in a variety of settings. A substantial challenge, however, is determining how to properly incorporate complicated outcome $Y$ into imputation models for missing covariates $X$, since the way in which the outcome is incorporated can have substantial impact on downstream analysis \citep{Beesley2016}. \citet{Bartlett2014} developed an imputation strategy that directly uses the target analysis model structure (e.g. $f(Y \vert X)$) to impute missing covariate values. This approach is appealing since it ensures that the imputation and analysis models are compatible with respect to the assumed relationship between $Y$ and $X$. However, the approach in \citet{Bartlett2014} can often be challenging to implement in many practical data analysis strategies, and existing software (e.g. R package \textit{smcfcs}) is limited in the analysis models supported. \\
\indent In this paper, we propose a novel imputation and data analysis strategy that involves (1) imputing missing covariates \textit{ignoring} the outcome $Y$, (2) stacking the multiple imputations to form a single dataset, (3) augmenting the dataset with weights based on the assumed analysis model structure, $f(Y \vert X)$, and (4) analyzing the weighted, stacked data using a novel estimator for standard errors. This imputation strategy avoids the problem of incorporating $Y$ into covariate imputation models entirely, but it still can produce valid estimates for the analysis model parameters through the use of weights. Additionally, the covariate imputation and outcome modeling steps are separated in this data analysis pipeline, allowing these steps to be implemented independently by different analysts. \\
\indent A limitation of data analysis based on stacked multiple imputations in general is the lack of convenient estimators for corresponding standard errors. In this paper, we develop a novel approach for estimating standard errors for stacked multiple imputations in \ref{louis}. This estimator can be applied in our particular substantive model compatible imputation strategy, but it can also be applied for general data analysis of multiply imputed data as an alternative to Rubin's rules. An advantage of the proposed data analysis approach over separate analysis of the imputed datasets as in Rubin's rules is that we can easily impose restrictions in model estimates \textit{across} multiple imputations such as in analyses with variable selection \citep{Wood2008}. A disadvantage of this approach is that is requires calculation of the score and information matrices for a given parametric model. However, these can be easily calculated using existing software in R for many popular parametric models. Our proposed estimator can be easily implemented for several analysis models (e.g. generalized linear models, Cox proportional hazards models) using our R package \textit{StackImpute}. Additional work is needed to extend this estimator to the setting with penalized likelihood estimation, particularly when the penalty function is not differentiable. \\
\indent Overall, this paper proposes a novel imputation strategy that is compatible with the analysis model while maintaining the flexibility of chained equations imputation methods for imputing missing covariates. Additionally, we propose an estimator for calculating standard errors from stacked multiple imputations that can be applied in general imputation settings along with corresponding R software, ultimately making the stacked imputation strategy easier to apply in practical data analysis.

\section{Acknowledgments}
The authors cite the many investigators (listed in \citet{Beesley2016}) in the University of Michigan Head and Neck Specialized Program of Research Excellence for their contributions to patient recruitment, specimen collection, and study conduct. This research is partially supported by NIH grant CA129102.\\

\bibliographystyle{plainnat}
\bibliography{Bib}


%

\end{document}